\documentclass[10pt, conference]{IEEEtran}
\usepackage{theorem}
\usepackage{graphicx}
\usepackage{braket}
\usepackage[dvipsnames]{xcolor}
\newcommand{\qedsymbol}{\rule{2mm}{2mm}}
\usepackage{arydshln}

\newtheorem{theorem}{Theorem}

\newtheorem{remark}{Remark}
\newtheorem{lemma}{Lemma}
\newtheorem{example}{Example}

\usepackage{bm}
\usepackage{amsfonts}
\newcommand{\diag}{{\rm{diag}}}

\usepackage{xcolor}

\newcommand{\defeq}{\stackrel{\Delta}{=}}

\usepackage[utf8]{inputenc} 
\usepackage[T1]{fontenc}
\usepackage{url}
\usepackage{ifthen}
\usepackage{cite}
\usepackage[cmex10]{amsmath} 

\begin{document}

\title{Entanglement-Assisted Coding for Arbitrary Linear Computations Over a Quantum MAC}

\author{Lei Hu \quad Mohamed Nomeir \quad Alptug Aytekin \quad Yu Shi \quad Sennur Ulukus \quad Saikat Guha\\
    \normalsize Department of Electrical and Computer Engineering\\
    \normalsize University of Maryland, College Park, MD 20742 \\
    \normalsize \emph{leihu@umd.edu} \quad \emph{mnomeir@umd.edu} \quad \emph{aaytekin@umd.edu} \quad \emph{shiyu@umd.edu} \quad \emph{ulukus@umd.edu} \quad \emph{saikat@umd.edu}}

\maketitle

\begin{abstract}
    We study a linear computation problem over a quantum multiple access channel (LC-QMAC), where $S$ servers share an entangled state and separately store classical data streams $W_1,\cdots, W_S$ over a finite field $\mathbb{F}_d$. A user aims to compute $K$ linear combinations of these data streams, represented as $Y = \mathbf{V}_1 W_1 + \mathbf{V}_2 W_2 + \cdots + \mathbf{V}_S W_S \in \mathbb{F}_d^{K \times 1}$. To this end, each server encodes its classical information into its local quantum subsystem and transmits it to the user, who retrieves the desired computations via quantum measurements. In this work, we propose an achievable scheme for LC-QMAC based on the stabilizer formalism and the ideas from entanglement-assisted quantum error–correcting codes (EAQECC). Specifically, given any linear computation matrix, we construct a self-orthogonal matrix that can be implemented using the stabilizer formalism. Also, we apply precoding matrices to minimize the number of auxiliary qudits required. Our scheme achieves more computations per qudit, i.e., a higher computation rate, compared to the best-known methods in the literature, and attains the capacity in certain cases.
\end{abstract}

\section{Introduction}
Entanglement is a transformative resource that enhances classical communication \cite{shi2021entanglement, nielsen2010quantum, hsieh2008entanglement}. For example, in superdense coding, it doubles the classical bits sent per qubit \cite{werner2001all}. Beyond communication, entanglement also enables communication-efficient classical computations over a quantum multiple access channel (QMAC) \cite{song2020capacity, song2021capacity, yao_capacity_MAC, N_sum_box,aytekin2023quantum}. It has been shown that, when utilizing the maximally entangled state, certain tasks can be executed with the optimal communication gain, such as those satisfied with the strongly self-orthogonal (SSO) condition \cite{yao_capacity_MAC,N_sum_box}. However, for general linear computations over a QMAC (LC-QMAC), which support a wider range of applications \cite{dutta2016short, ramamoorthy2019universally, das2019distributed, yu2017polynomial}, the communication-efficient strategy is still unknown. This work aims to develop a coding scheme that achieves communication efficiency for LC-QMAC.

\begin{figure}[ht]
      \centering
      \includegraphics[width=0.39\textwidth]{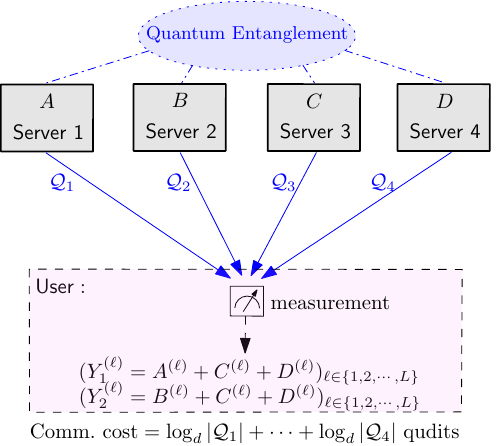}
      \caption{An example of LC-QMAC with $S=4$ servers and data streams $A, B, C, D$. The user wants $K=2$ linear combinations of the data streams.}
      \label{fig:model}
\end{figure}

Fig.~\ref{fig:model} shows an example of LC-QMAC, where $S=4$ servers separately store the data streams $A,B,C,D$, and an entangled state is shared among the servers. Meanwhile, a user aims to compute $K=2$ linear combinations: $Y_1 = A + C + D$ and $Y_2 = B + C + D$. To achieve this, each server $s$ encodes its classical information into its quantum subsystem $\mathcal{Q}_s$, which is transmitted to the user through a noise-free QMAC. Upon receiving all quantum subsystems $\mathcal{Q}_1\mathcal{Q}_2\mathcal{Q}_3 \mathcal{Q}_4$, the user performs quantum measurements to recover $L$ instances of the computations: $Y_1^{(\ell)} = A^{(\ell)} + C^{(\ell)} + D^{(\ell)}, \ Y_2^{(\ell)} = B^{(\ell)} + C^{(\ell)} + D^{(\ell)}, \ \ell \in [L]$. The normalized download cost tuple is defined as $(\log_d |\mathcal{Q}_1|/L, \log_d |\mathcal{Q}_2|/L, \cdots, \log_d |\mathcal{Q}_4|/L)$, where $|\mathcal{Q}_s|, \ s \in [4],$ is the dimension of $\mathcal{Q}_s$, and $\log_d |\mathcal{Q}_s|$ represents the number of downloaded qudits from server $s$. A cost tuple is called achievable if the user can recover the desired computations using that tuple. A region is achievable if all tuples within it are achievable. The computation rate is defined as the number of computation dits acquired per transmitted qudit, i.e., $2L/\sum_{s = 1}^{4} \log_d |\mathcal{Q}_s|$.
 
Prior work \cite{yao_capacity_MAC} studied the $\Sigma$-QMAC model, where the user typically requires a single sum-computation of the data streams. The optimal achievable scheme for $\Sigma$-QMAC is constructed via the SSO matrix. The key distinction between LC-QMAC and $\Sigma$-QMAC lies in the user's requirements: in LC-QMAC, the user seeks multiple computations instead of just one. Applying the $\Sigma$-QMAC scheme to LC-QMAC by retrieving each computation separately may result in a significantly high communication cost compared to the optimal approach. In fact, retrieving multiple computations together requires fewer qudits than retrieving each computation individually. Additionally, for the LC-QMAC model, prior work \cite{yao2024inverted} provided lower bounds on the minimum number of downloaded qudits from the servers in LC-QMAC. However, a general achievable scheme and the corresponding achievable computation rate remain unaddressed.

The motivation of this work is to develop an achievable scheme that offers insights into implementing linear computations over a QMAC. The central question is: given an arbitrary linear computation matrix, how can we design a quantum coding scheme that fully leverages the communication gains enabled by multiparty entanglement? Alternatively, since certain computation structures allow for maximum communication efficiency, how can we expand the desired computation matrix into a large space to identify a feasible structure -- potentially at the cost of downloading additional qudits?

Our approach to LC-QMAC builds on the stabilizer formalism and the ideas from entanglement-assisted quantum error–correcting codes (EAQECC) \cite{TIT_catalytic,science_brun}. In a QMAC, the stabilizer formalism ensures that a linear computation can be implemented if the computation matrix satisfies the self-orthogonal (SO) condition \cite{gottesman1997stabilizer}. The challenge, however, is to design an SO matrix for an arbitrary linear computation matrix. Inspired by EAQECC, which utilize preshared entangled pairs to construct a stabilizer group from a set of noncommuting generators by enlarging these generators, we adapt this technique for LC-QMAC. Specifically, servers transmit auxiliary entangled qudits, expanding the linear computation matrix to construct an SO matrix. To further reduce the number of auxiliary qudits, precoding matrices are applied and optimized to restructure the computation matrix without altering the computation results.

Let us return to the example in Fig.~\ref{fig:model}. We evaluate two natural baselines for comparison. In the classical setting where no quantum resources are available to the servers, the user must download all $4$ dits $(A, B, C, D)$ to obtain the two computations $Y_1$ and $Y_2$. This results in a rate of $1/2$ computations/dit. The second baseline is to retrieve $Y_1$ and $Y_2$ separately by applying the method described in $\Sigma$-QMAC \cite{yao_capacity_MAC}. This approach requires $3$ qudits communication cost, yielding a rate of $2/3$ computations/qudit. In contrast, the proposed method in this work achieves the communication cost of $2.5$ qudits, resulting in an improved rate of $4/5$ computations/qudit.

\textit{Notation:}
For integer $M$, $[M]$ denotes the set $\{1, \cdots, M\}$. For integers $a$ and $b$, $[a:b]$ denotes  $\{a,\cdots,b \}$. $A_{[M]}$ and $A^{[L]}$ are the compact notations of $\{A_{1}, A_2, \cdots, A_{M}\}$ and $\{A^{(1)}, A^{(2)}, \cdots, A^{(L)}\}$, respectively. $\mathbb{F}_{d}$ denotes a finite field with order $d$, where $d = p^r$ with a prime number $p$ and a positive integer $r$. $\mathbb{F}_{d}^{a \times b}$ is the set of $a \times b$ matrices with elements in $\mathbb{F}_{d}$. $\mathrm{blkdiag}(\cdot)$ is a block matrix where all off-diagonal blocks are zero matrices. $\mathbf{I}_{a}$ is an $a \times a$ identity matrix, and $\mathbf{0}_{a \times b}$ is an $a \times b$ matrix with all zero elements. $\otimes$ denotes the tensor product. Empty sub-blocks of matrices are zeros. For a $d$-dimensional quantum system, the computational basis is denoted as $\{ \ket{j} \}_{j \in \mathbb{F}_d}$. The operators $\mathsf{X}(x)$ and $\mathsf{Z}(z)$ are defined as $\mathsf{X}(x)\ket{j} =  \ket{j+x}$ and $\mathsf{Z}(z) \ket{j} = \omega^{{\rm{tr}}(jz)} \ket{j}$ for $j,x,z \in \mathbb{F}_d$, where $\omega \defeq e^{2\pi i /p}$ and ${\rm{tr}}(x) \defeq \sum_{j=0}^{r-1} x^{p^j} \in \mathbb{F}_p$.

\section{Problem Formulation}
\subsection{System Model}
An LC-QMAC problem is characterized by the linear combination matrices $\mathbf{V}_{[S]}$ over $\mathbb{F}_d$. Specifically, we consider a QMAC consisting of $S$ distributed servers and a user. The servers share a quantum entanglement and store classical data streams. Each server $s, \ s \in [S],$ stores $L$ instances of the data stream $W_s^{[L]} = [W_s^{(1)},W_s^{(2)},\cdots,W_s^{(L)}]$, where $W_s^{(\ell)}\in \mathbb{F}_d^{m_s \times 1}$ is the $\ell$th instance. For each instance, the user aims to compute $K$ linear combinations of the data streams, given by
\begin{align}
    Y^{(\ell)} & = \mathbf{V}_1 W_1^{(\ell)} + \mathbf{V}_2 W_2^{(\ell)} + \cdots + \mathbf{V}_S W_S^{(\ell)}, \quad \ell \in [L], \label{eq:formula_Y}
\end{align}
where $\mathbf{V}_s \in \mathbb{F}_d^{K \times m_s}$ is the linear combination matrix associated with $W_s$. 
Defining $\mathbf{V} \defeq [\mathbf{V}_1, \mathbf{V}_2, \cdots, \mathbf{V}_S]$ as the linear computation (LC) matrix and $W^{(\ell)} \defeq [(W_1^{(\ell)})^T, (W_2^{(\ell)})^T, \cdots, (W_S^{(\ell)})^T]^T$, (\ref{eq:formula_Y}) is equivalent to $Y^{(\ell)} = \mathbf{V} W^{(\ell)}$. Without loss of generality, we assume that: 
\begin{enumerate}
    \item $\mathrm{rank}(\mathbf{V}) = K$, ensuring that the desired computations are linearly independent combinations of the data streams;

    \item $\mathrm{rank}(\mathbf{V}_s) = m_s$ and $m_s \leq  K \leq \sum_{s \in [S]}m_s, \ \forall s \in [S]$, which guarantee that $\mathbf{V}_s W_s^{(\ell)}$ preserves all information in $W_s^{(\ell)}$ without redundancy.
\end{enumerate}

For all $L$ instances, the computations can be written compactly as
\begin{align}
    Y^{[L]} = \mathbf{V}_1 W_1^{[L]} + \mathbf{V}_2 W_2^{[L]} + \cdots + \mathbf{V}_S W_S^{[L]},\label{eq:Y_L}
\end{align}
where $Y^{[L]} \in \mathbb{F}_d^{K \times L}$ represents the concatenation of all $L$ instances of computations.

To compute $Y^{[L]}$, the servers encode the classical information into the quantum system and transmit the system to the user. The user retrieves the desired computations through quantum measurements. The details of the process are described below.

\subsection{Communication Scheme}
Given $\mathbf{V}_{[S]}$, the communication process of the LC-QMAC is composed of three main steps:
\begin{enumerate}
    \item \textbf{(Entanglement Distribution)}
    Each server $s \in [S]$ possesses a quantum subsystem $\mathcal{Q}_s$, and the servers share a globally entangled quantum state $\mathcal{Q} = \mathcal{Q}_1 \cdots \mathcal{Q}_S$ across these subsystems. This entangled state, denoted by the density matrix $\rho^{\rm ini}$, is pre-established before the communication begins. The dimension of each quantum subsystem $\mathcal{Q}_s$ is given by $\delta_s \defeq |\mathcal{Q}_s|$, and the number of qudits is $\log_d \delta_s$.
    
    \item \textbf{(Information Encoding)}
    Each server $s\in [S]$ performs a local operation on its quantum subsystem $\mathcal{Q}_s$ to encode classical information into the quantum system. Specifically, server $s$ applies a unitary matrix $U_s$, which is a function of the classical data stream $W_s$ at server $s$. After this encoding operation, the density matrix of the quantum state becomes $\rho^{\rm enc} = (U_1 \otimes U_2 \otimes \cdots \otimes U_S) \rho^{\rm ini} (U_1^\dag \otimes U_2^\dag \otimes \cdots \otimes U_S^\dag)$. Once the encoding is complete, all servers send their quantum subsystems to the user through a noiseless QMAC.

    \item \textbf{(Computations Decoding)}
    Upon receiving the composite quantum system $\mathcal{Q}_1\mathcal{Q}_2 \cdots \mathcal{Q}_S$, the user performs a positive-operator-valued measure (POVM) and gets the measurement outcomes. These outcomes ensure that the user can successfully retrieve the desired computations $Y^{[L]}$ in (\ref{eq:Y_L}).
\end{enumerate}

\subsection{Achievable Region}
Based on the communication process, we now define the metrics used to evaluate the communication efficiency of the LC-QMAC.
Given a communication scheme, let us define the \textit{normalized download cost tuple} as
\begin{align}
    \bm{\Delta} = (\Delta_1, \Delta_2, \cdots, \Delta_S) \defeq \left(\frac
    {\log_d \delta_1}{L}, \frac
    {\log_d \delta_2}{L}, \cdots, \frac
    {\log_d \delta_S}{L} \right).
\end{align}
A download cost tuple $\bm{\Delta}$ is called \textit{achievable} if there exists a communication scheme and $L$ such that the user can successfully retrieve $L$ instances of the computations $Y^{[L]}$ by downloading $L \Delta_s$ qudits from each server $s$. Furthermore, a region $\mathcal{D}_{\mathsf {LC}}$ is called achievable if every tuple $ \bm{\Delta} \in \mathcal{D}_{\mathsf{LC}}$ is achievable.
The \textit{computation rate} $R_{\mathsf {LC}}$ is defined as 
\begin{align}
    R_{\mathsf {LC}}
     \defeq \frac{KL}{ \sum_{s\in [S]} \log_d\delta_{s}  } = \frac{K}{ \sum_{s \in [S]} \Delta_{s}}.
\end{align}
A rate $R_{\mathsf {LC}}$ is called achievable if $\bm{\Delta} \in \mathcal{D}_{\mathsf {LC}}$.
Finally, the capacity $C_{\mathsf{LC}} $ is defined as the supremum of all achievable computation rates.

\subsection{An Implementation Based on the Stabilizer Formalism}
The stabilizer formalism establishes that an LC matrix satisfying the SO condition can be effectively implemented in a QMAC system. This key insight is summarized as follows:
\begin{lemma}\label{lemma-SO}
    (Stabilizer-Based Implementation \cite{N_sum_box,ketkar2006nonbinary,ashikhmin2001nonbinary,calderbank1998quantum}) 
    Consider a distributed system with $N$ parties and a receiver, where each party $n \in [N]$ has a qudit and classical symbols $(x_n, x_{n+N}) \in \mathbb{F}_d^2$. The parties share a quantum system $\mathcal{Q} = \mathcal{Q}_1 \cdots \mathcal{Q}_N$. Each party $n $ applies the Pauli operator ${\mathsf X}(x_n) \mathsf{Z}(x_{n+N})$ to their respective subsystem $\mathcal{Q}_n$. This encoding defines a unitary matrix $\Tilde{\mathbf{W}}(\mathbf{x}) \defeq \mathsf{X}(x_1)\mathsf{Z}(x_{1+N}) \otimes \cdots \otimes \mathsf{X}(x_N)\mathsf{Z}(x_{2N})$ where $\mathbf{x} \defeq [x_1, \cdots, x_{2N}]^T$. Then, given an SO subspace $\mathcal{V}$ in $\mathbb{F}_d^{2N}$, there exists a stabilizer group defined as
    \begin{align}
        \mathcal{L}(\mathcal{V}) \defeq \{c_\mathbf{v} \Tilde{\mathbf{W}}(\mathbf{v}) \lvert c_\mathbf{v} \in \mathbb{C}, \ \mathbf{v} \in \mathcal{V}\},
    \end{align}
    where $c_\mathbf{v} \mathbf{I}_{d^N} \in \mathcal{L}(\mathcal{V})$ if and only if $c_\mathbf{v} = 1$. The quotient space of $\mathcal{V}$ defines a set of projective value measures (PVMs) that the receiver can apply to extract the measurement outcome $\mathbf{y} \in \mathbb{F}_d^\kappa$, $\kappa \leq N$. Based on this, the input-output relationship between the $N$ parties and the receiver is given by $\mathbf{y} = \mathbf{M} \mathbf{x}$, where the transfer matrix $\mathbf{M} \defeq [\mathbf{M}_l, \mathbf{M}_r]$ with $\mathbf{M}_l, \mathbf{M}_r \in \mathbb{F}_d^{\kappa \times N}$ satisfying the SO property: $\mathrm{rank}([\mathbf{M}_l, \mathbf{M}_r]) = \kappa$ and $\mathbf{M}_r \mathbf{M}_l^T = \mathbf{M}_l \mathbf{M}_r^T$.
\end{lemma}
        
However, a general LC matrix $\mathbf{V}$ does not naturally satisfy the SO property. To overcome this limitation, we propose a new method to construct an SO matrix for any given $\mathbf{V}$, thereby formulating a new achievable region for LC-QMAC.

\section{Main Results}
\begin{theorem}\label{thm-1}
    Given any linear combination matrices $\mathbf{V}_{[S]}$, the following download cost region is achievable,
    \begin{align}
        \mathcal{D}_{\mathsf{LC}}=\left\{\bm{\Delta} \in \mathbb{R}_{+}^S  \left\lvert\, 
        \begin{aligned} 
            &  2 \Delta_s \geq m_s, \ s \in [S], \\ 
            & \sum_{s \in [S]}  2\Delta_s \geq \sum_{s \in [S]} m_s+ c
        \end{aligned}\right.\right\},
    \end{align}
    where 
    \begin{align}
        c = \min_{ \det(\mathbf{P}_s) \neq 0,\ s \in [S] } \mathrm{rank}\left(\sum_{s \in [S]} \mathbf{V}_s \mathbf{P}_s \mathbf{V}_s^T \right),
    \end{align}
    and $\mathbf{P}_s$ is the precoding matrix at server $s$.
    Consequently, the computation rate 
    \begin{align}
        R_{\mathsf{LC}} = \frac{2K}{\sum_{s \in [S] } m_s + c}
    \end{align}
    is achievable.
\end{theorem}

The proof is given in Section \ref{Sec:achievable_scheme}.

\begin{remark}
     The key idea of the proposed scheme to achieve $\mathcal{D}_{\mathsf{LC}}$ is for the user to download auxiliary qudits from the servers, therefore expanding the LC matrix and enabling the construction of an SO matrix.
     Specifically, consider $L=2$ instances, where the computation task becomes
     \begin{align}
        \begin{bmatrix}
            Y^{(1)} \\
            Y^{(2)}
        \end{bmatrix}
        =
        \left[\begin{array}{c:c}
                    \mathbf{V} & \\
                    &  \mathbf{V}
              \end{array}\right]
        \begin{bmatrix}
            W^{(1)} \\
            W^{(2)}
        \end{bmatrix}. \label{eq:instance2}
    \end{align}
    Note that the transfer matrix $\mathbf{M} \defeq \mathrm{blkdiag}(\mathbf{V}, \mathbf{V})$ in (\ref{eq:instance2}) may not satisfy the SO condition. To address this, the servers totally transmit $c$ auxiliary entangled qudits, with no operations performed on these qudits (i.e., the encoded classical information is $0$). These auxiliary qudits add $c$ columns to both the left and right halves of $\mathbf{M}$. By carefully designing these columns, the modified matrix can be ensured to meet the SO condition. In terms of the communication cost, each server $s \in [S]$ sends $m_s$ qudits with encoded classical information, while all servers collectively transmit $c$ auxiliary qudits to the user. As a result, a total of $\sum_{s \in [S]} m_s + c$ qudits are transmitted, enabling the user to acquire $2K$ computation symbols.
\end{remark}
 
\begin{remark}
    From (\ref{eq:instance2}), the value of $c$ depends on the structure of the LC matrix $\mathbf{V}$.
    Furthermore, $c$ can be reduced if servers apply precoding matrices to their data streams.
    Specifically, if each server $s \in [S]$ applies $\mathbf{P}_s^{-1} \in \mathbb{F}_d^{m_s \times m_s}$ to its data stream $W_s^{(1)}$, the computation task becomes
    \begin{align}
        \begin{bmatrix}
            Y^{(1)} \\
            Y^{(2)}
        \end{bmatrix}
        =
        \left[\begin{array}{c:c}
                    \mathbf{V} \mathbf{P} & \\
                    &  \mathbf{V}
              \end{array}\right]
        \begin{bmatrix}
            \mathbf{P}^{-1}W^{(1)} \\
            W^{(2)}
        \end{bmatrix}, \label{eq:VP}
    \end{align}
    where $ \mathbf{P} = \mathrm{blkdiag}(\mathbf{P}_1, \cdots, \mathbf{P}_S)$. Now, $\mathrm{blkdiag}(\mathbf{V}\mathbf{P}, \mathbf{V})$ in (\ref{eq:VP}) is the effective LC matrix with precoding. This allows us to design the precoding matrices $\mathbf{P}_{[S]}$ to minimize $c$, as long as each $\mathbf{P}_s$ is invertible.\footnote{We will show that applying precoding to two instances results in the same communication cost as applying it to only one instance.}
\end{remark}

\begin{remark}
    As a special case, for the summation problem  $\Sigma$-QMAC studied in \cite{yao_capacity_MAC}, the LC matrix is  $\mathbf{V} = [1,1,\cdots,1]$ and $K=1$. Then, the value of $c$ in Theorem~\ref{thm-1} is
    \begin{align}
        c = \min_{p_s \neq 0, \ s \in [S]} \mathrm{rank}(\sum_{s \in [S]} p_s ) = 0,
    \end{align}
    where the last equality holds when $\sum_{s \in [S]} p_s = 0$.
    Consequently, the rate $R_{\mathsf{LC}} = 2/S$ is achievable, which coincides with the capacity of $\Sigma$-QMAC as established in \cite{yao_capacity_MAC}.
    This demonstrates that the proposed scheme is a generalization of the method in $\Sigma$-QMAC and achieves the capacity for this case.
\end{remark}
        
\section{Illustrative Examples}
We now present examples to show the construction of SO matrices and demonstrate how the proposed approach achieves higher rates compared to existing schemes. Notably, the capacities of LC-QMAC can be achieved in certain cases.

\begin{example}\label{example1}
Let us go back to the example in Fig.~\ref{fig:model}, where the LC matrix is given by
\begin{align}
    \mathbf{V}= \left[\begin{array}{c:c:c:c}
        1 & 0 & 1 & 1 \\
        0 & 1 & 1 & 1
    \end{array}
    \right]
\end{align}
in $\mathbb{F}_d$.
Note that this problem involves two sums. If each is computed separately using the method in \cite{yao_capacity_MAC}, then each requires $1.5$ qudits. This leads to an achievable rate of $2/3$ computations/qudit. In contrast, by Theorem~\ref{thm-1}, the number of required auxiliary qudits is
\begin{align}
    c & = \min_{p_1, \cdots, p_4 \neq 0} \mathrm{rank}\left(
    \begin{bmatrix}
        p_1 + p_3 + p_4 & p_3 + p_4 \\
        p_3 + p_4 & p_2 + p_3 + p_4
    \end{bmatrix}
    \right) \\
    & \geq \min_{p_3, p_4 \neq 0} 2 - \mathrm{rank}\left(
    \begin{bmatrix}
        p_3  + p_4 &  p_3  + p_4\\
        p_3  + p_4 & p_3  + p_4
    \end{bmatrix}
    \right) \label{eq:ineq_rank} \\
    & \geq 1,
\end{align}
where (\ref{eq:ineq_rank}) holds because $\mathrm{rank}(\mathbf{A} + \mathbf{B}) \geq |\mathrm{rank}(\mathbf{A}) - \mathrm{rank}( \mathbf{B})| $ and $p_1, p_2 \neq 0$.  
Also, $c = 1$ when $p_1 = p_3 + p_4$,  $2p_2 + p_3 + p_4 = 0$, and $d\geq 3$. 
For example, when $d=3$, a valid SO matrix can be constructed as
\begin{align}
    \mathbf{M} = 
    \left[\begin{array}{ccccc:ccccc}
        2 & 0 & 1 & 1 & \color{Brown}{2} & & & & \\
        0 & 2 & 1 & 1 & \color{Brown}{1} & & & & \\
        & & & & & 1 & 0& 1 & 1 & \color{Brown}{1}\\
        & & & & & 0 & 1& 1 & 1 & \color{Brown}{2}
    \end{array}\right],
\end{align}
which satisfies the SO condition in $\mathbb{F}_3$.
Using this approach, the total number of qudits required for $L=2$ instances is $4+1=5$, resulting in a rate of $4/5$ computations/qudit (each instance has $2$ computations).
This demonstrates the communication efficiency of the proposed scheme in reducing the total communication cost while jointly retrieving computation symbols.
\end{example}

\begin{example}\label{example:sum}
 Consider the same setting as in Example~\ref{example1}, but now the user aims to compute $Y_1 = W_1 + W_3$ and $ Y_2 = W_2+ W_4$ in $\mathbb{F}_d$. The corresponding LC matrix is given by
\begin{align}
    \mathbf{V} = \left[ \begin{array}{c:c:c:c}
        1 & 0 & 1 & 0 \\
        0 & 1 & 0 & 1
    \end{array} \right].
\end{align}
Using Theorem~\ref{thm-1}, the number of auxiliary qudits required to construct an SO matrix is calculated as
\begin{align}
    c = \min_{p_1, \cdots, p_4 \neq 0} \mathrm{rank}\left( 
    \begin{bmatrix}
        p_1 + p_3 & 0 \\
        0 & p_2 + p_4
    \end{bmatrix}
    \right) = 0,
\end{align}
where the second equality holds by setting $p_1 + p_3 = 0$ and $p_2 + p_4 = 0$.
This result indicates that precoding matrices suffice to construct an SO matrix, and no additional qudits are needed. For instance, by letting $p_1 = p_2 = d-1$ and $p_3 = p_4 = 1$, the SO matrix is designed as
\begin{align}
    \mathbf{M} = 
    \left[
    \begin{array}{cccc:cccc}
        d-1 & 0 & 1 & 0 &  &  &  &  \\
        0 & d-1 & 0 & 1 &  &  &  &  \\
         &  &  &  & 1 & 0 & 1 & 0 \\
         &  &  &  & 0 & 1 & 0 & 1
    \end{array}
    \right].
\end{align}
This construction achieves a total download cost of $4$ qudits for $L=2$ instances. Thus, the achievable rate is $1$ computation/qudit, which is the capacity by Holevo bound \cite{holevo1973bounds}.
\end{example}

\begin{remark}
    In Example~\ref{example:sum}, when no precoding is applied, i.e., $p_i = 1, \ i \in [4]$, the required number of auxiliary qudits is $c = \mathrm{rank} (\mathbf{V} \mathbf{V}^T) = 2 $ if $d \neq 2$. This means precoding can help reduce the communication cost for larger $d$.
\end{remark}

\begin{example}
Consider the setting where
$\mathbf{V}_1 = \mathbf{I}_{S}$ and $ \mathbf{V}_s = \mathbf{e}_1$, for all $s \in [2:S]$, where $\mathbf{e}_1 \defeq [1, 0,0, \cdots, 0]^T$.
Applying Theorem~\ref{thm-1}, the number of auxiliary qudits for $L = 2$ instances is
\begin{align}
    c & = \min\limits_{\substack{\det(\mathbf{P}_1) \neq 0, \\ p_s \neq 0, \ s \in [2:S]}}
     \mathrm{rank}(\mathbf{P}_1 + \diag(\sum_{s \in [2:S]} p_s, 0, \cdots, 0)) \\
     &\geq \min\limits_{\substack{\det(\mathbf{P}_1) \neq 0, \\ p_s \neq 0, \ s \in [2:S]}} S - \mathrm{rank} (\diag(\sum_{s \in [2:S]} p_s, 0, \cdots, 0)) \\
     & \geq S - 1,
\end{align}
where $c = S-1$ holds by setting  $\mathbf{P}_1 = \diag(p_{11}, p_{22}, \cdots, p_{SS})$ with $p_{11}, p_{22}, \cdots,p_{SS} \neq 0$, and $p_{11} + \sum_{s \in [2:S]}p_s = 0$.
The computation rate achieved by the proposed approach is then given by
\begin{align}
    R_{\mathsf{LC}} = \frac{2S}{2S-1 + c} = \frac
    {S}{3S/2-1}.
\end{align}
It was shown in \cite{yao2024inverted} that $C_{\mathsf{LC}} \leq \frac{S}{3S/2 -1}$. The proposed approach achieves this rate, confirming that the capacity in this case is $ C_{\mathsf{LC}} = \frac{S}{3S/2 -1}$.
\end{example}
    
\section{The General Achievable Scheme for LC-QMAC}\label{Sec:achievable_scheme}
The general scheme is based on $L=2$ instances. The computation task is to implement the transfer matrix $\mathbf{M} = \mathrm{blkdiag}(\mathbf{V}, \mathbf{V})$ in (\ref{eq:instance2}).
The SO condition requires that $\mathbf{V} \mathbf{V}^T = \mathbf{0}_{K \times K}$, since $\mathbf{M}$ already satisfies $\mathrm{rank}(\mathbf{M}) = 2K $. However, this condition is not generally satisfied by arbitrary $\mathbf{V}$.
To mitigate this issue, we first introduce a local precoding matrix at each server, such that the computation task becomes
\begin{align}
    \begin{bmatrix}
        Y^{(1)} \\
        Y^{(2)}
    \end{bmatrix}= \underbrace{\left[\begin{array}{c:c}
        \mathbf{V}\mathbf{P}^{(1)} & \\
        &  \mathbf{V}\mathbf{P}^{(2)}
    \end{array}\right]}_{\defeq \ \bar{\mathbf{M}}}
    \begin{bmatrix}
        (\mathbf{P}^{(1)})^{-1} W^{(1)} \\
        (\mathbf{P}^{(2)})^{-1}W^{(2)}
    \end{bmatrix},\label{eq:M_bar}
\end{align}
where $\mathbf{P}^{(\ell)} = \mathrm{blkdiag}(\mathbf{P}^{(\ell)}_1, \cdots, \mathbf{P}^{(\ell)}_S), \ \ell =1,2$, and each $\mathbf{P}^{(\ell)}_s$ is an invertible matrix. Thus, $(\mathbf{P}^{(\ell)})^{-1} = \mathrm{blkdiag}((\mathbf{P}^{(\ell)}_1)^{-1}, (\mathbf{P}_2^{(\ell)})^{-1}, \cdots, (\mathbf{P}^{(\ell)}_S)^{-1})$. Here, $(\mathbf{P}^{(\ell)}_s)^{-1}$ is the precoding matrix that server $s$ applies to its  data stream $W_s^{(\ell)}$.
The introduction of the precoding matrix provides additional flexibility in implementing the transfer matrix $\bar{\mathbf{M}}$ in (\ref{eq:M_bar}).

Now, the task is to implement the $\bar{\mathbf{M}}$. 
Inspired by the theory of EAQECC, we leverage auxiliary qudits to enlarge the transfer matrix to construct an SO matrix. Specifically, each server $s \in [S]$ transmits $m_s$ qudits carrying encoded classical information, while all servers collectively transmit $c$ auxiliary qudits without performing any operations. The user can get
\begin{align}
\begin{bmatrix}
    Y^{(1)} \\
    Y^{(2)}
\end{bmatrix} =
\underbrace{\left[
\begin{array}{cc:cc}
    \mathbf{V} \mathbf{P}^{(1)} & \color{Brown}{\bar{\mathbf{V}}'} &  & \\
    & & \mathbf{V}\mathbf{P}^{(2)} & \color{Brown}{\mathbf{V}'}
\end{array}
\right]}_{\defeq \ \bar{\mathbf{M}}'}
\begin{bmatrix}
    (\mathbf{P}^{(1)})^{-1} W^{(1)} \\ \color{Brown}{\mathbf{0}_{c \times 1}} \\ \hdashline
    (\mathbf{P}^{(2)})^{-1} W^{(2)} \\
    \color{Brown}{\mathbf{0}_{c \times 1}}
\end{bmatrix},\label{eq:M'}
\end{align}
where $\bar{\mathbf{V}}'$ and $\mathbf{V}'$ are designed such that the $\bar{\mathbf{M}}'$  satisfies the SO condition

Next, we present the following Lemmas to show the closed-form expression for $c$ and the existence of $\bar{\mathbf{V}}'$ and ${\mathbf{V}}'$.

\begin{lemma}\label{lemma:symplectic}
    Consider a block diagonal matrix 
    \begin{align}
    \mathbf{M}_o \defeq 
    \left[\begin{array}{c:c}
        \mathbf{H} & \\
        & \mathbf{G}
    \end{array}\right],\label{eq:M_o}
    \end{align}
    where $\mathbf{H} = [\mathbf{h}_1, \cdots, \mathbf{h}_K]^T \in \mathbb{F}_d^{K \times m}$, $ \mathbf{G} = [\mathbf{g}_1, \cdots, \mathbf{g}_K]^T \in  \mathbb{F}_d^{K \times m}$ with $K \leq m$ and $\mathrm{rank}(\mathbf{H}) = \mathrm{rank}(\mathbf{G}) = K$. There exist a nonnegative integer $c$ and invertible matrices $\mathbf{B}_1, \mathbf{B}_2 \in \mathbb{F}_d^{K \times K}$ such that the transformed matrices $\bar{\mathbf{H}} = \mathbf{B}_1 \mathbf{H} $ and $\bar{\mathbf{G}} = \mathbf{B}_2 \mathbf{G}  $ satisfy the following properties:
    \begin{align}
        \bar{\mathbf{h}}_i^T \bar{\mathbf{g}}_i & \neq 0, \quad i \in [c], \label{eq:property-1}\\
        \bar{\mathbf{h}}_i^T \bar{\mathbf{g}}_j & = 0, \quad \forall i \neq j, \\
        \bar{\mathbf{h}}_i^T \bar{\mathbf{g}}_i & = 0, \quad i \in [c:K], \label{eq:property-4}
    \end{align}
    where $\bar{\mathbf{H}} = [\bar{\mathbf{h}}_1, \cdots, \bar{\mathbf{h}}_K]^T$ and $\bar{\mathbf{G}} = [\bar{\mathbf{g}}_1, \cdots, \bar{\mathbf{g}}_K]^T$.
\end{lemma}

The proof is presented in Appendix~\ref{proof:lemma_symplectic}.

\begin{remark}
    Lemma~\ref{lemma:symplectic} establishes that, given the matrices $\mathbf{H}$ and $\mathbf{G}$ in (\ref{eq:M_o}), they can be transformed into $\bar{\mathbf{H}}$ and $\bar{\mathbf{G}}$, where there are $c$ pairs of vectors $(\bar{\mathbf{h}}_i, \bar{\mathbf{g}}_i)$ that are not orthogonal.
    This implies that by adding $c$ columns to both $\bar{\mathbf{H}}$ and $\bar{\mathbf{G}}$, the resulting enlarged matrices can be made orthogonal. Moreover, by leveraging the transformation matrices $\mathbf{B}_1$ and $\mathbf{B}_2$, we can add $c$ columns to $\mathbf{H}$ and $\mathbf{G}$ to ensure their orthogonality.
\end{remark}

\begin{lemma}\label{lemma:c}
    For the matrix $\mathbf{M}_o$ defined in (\ref{eq:M_o}), there exist matrices $\mathbf{H}', \mathbf{G}' \in \mathbb{F}_d^{K \times c}$ such that the expanded matrix
    \begin{align}
    \mathbf{M}_o' = 
    \left[
        \begin{array}{cc:cc}
            \mathbf{H} & \color{Brown}{\mathbf{H}'} & & \\
            & & \mathbf{G} & \color{Brown}{\mathbf{G}'}
        \end{array} \label{eq:designed_M}
        \right]
    \end{align}
    is an SO matrix, and $c = \mathrm{rank}(\mathbf{H} \mathbf{G}^T)$.
\end{lemma}

The proof is presented in Appendix~\ref{proof:lemma_c}.

Based on Lemma~\ref{lemma:symplectic} and Lemma~\ref{lemma:c}, the
$\bar{\mathbf{V}}'$ and ${\mathbf{V}}'$ in (\ref{eq:M'}) can be designed to construct an SO matrix $\bar{\mathbf{M}}'$.
Also, the required number of auxiliary qudits is $c = \mathrm{rank}(\mathbf{V} \mathbf{P}^{(1)}(\mathbf{P}^{(2)})^T \mathbf{V}^T) = \mathrm{rank}\left(\sum_{s \in [S]} \mathbf{V}_s \mathbf{P}^{(1)}_s (\mathbf{P}^{(2)}_s)^T \mathbf{V}_s^T \right)$.
Since the precoding matrices can be optimized, $c$ can be minimized over all invertible matrices $\mathbf{P}_{[S]}^{(1)}$ and $\mathbf{P}_{[S]}^{(2)}$. The minimum $c$ is equivalent to $\min_{ \det(\mathbf{P}_s) \neq 0,\ s \in [S] } \mathrm{rank}(\sum_{s \in [S]} \mathbf{V}_s \mathbf{P}_s \mathbf{V}_s^T )$ when setting $\mathbf{P}_s^{(1)}(\mathbf{P}_s^{(2)})^T =\mathbf{P}_s, \ s \in [S]$.
After determining the optimal $\mathbf{P}_s^{*}$, the precoding matrices $\mathbf{P}_s^{(1)} $ and $\mathbf{P}_s^{(2)}$ are designed as $\mathbf{P}_s^{*}$ and $\mathbf{I}_{ m_s}$, respectively.
This implies that precoding for one instance achieves the same communication cost as precoding for two instances in the proposed approach.

As a result, if the number of downloaded qudits satisfies
\begin{align}
    \log_d \delta_s = m_s,\ \forall s \in  [S],\   \sum_{s \in [S]} \log_d \delta_s = \sum_{s \in [S]} m_s + c, \label{eq:region_L}
\end{align}
the user can recover $L = 2$ instances of computations. By normalizing (\ref{eq:region_L}) over $L$, we have $ \Delta_s =  \log_d \delta_s/L = m_s/{2},\ \forall s \in  [S] $ and $
\sum_{s \in [S]} \Delta_s= \sum_{s \in [S]} \log_d \delta_s/L= (\sum_{s \in [S]} m_s + c)/2$, which means the region in Theorem~\ref{thm-1} is achievable.

\appendix 

\subsection{Proof of Lemma~\ref{lemma:symplectic}} \label{proof:lemma_symplectic}
By Smith normal form \cite{newman1997smith}, there exist \textit{invertible} matrices $\mathbf{U}_1$, $\mathbf{U}_2 \in \mathbb{F}_d^{K \times K}$ such that
\begin{align}
    \mathbf{U}_1  \mathbf{H} \mathbf{G}^T \mathbf{U}_2 = \mathbf{\Lambda},
\end{align}
where $\mathbf{\Lambda}$ is a diagonal matrix with the first $c$ diagonal elements being nonzero.
Define $\mathbf{B}_1 = \mathbf{U}_1$ and $\mathbf{B}_2 = \mathbf{U}_2^T$. Then, let 
$\bar{\mathbf{H}} = \mathbf{B}_1 \mathbf{H}$ and $\bar{\mathbf{G}} = \mathbf{B}_2 \mathbf{G}$. We obtain 
\begin{align}
    \bar{\mathbf{H}} \bar{\mathbf{G}}^T =
    \left[\begin{array}{c:c}
        \diag( \bar{\mathbf{h}}_1^T \bar{\mathbf{g}}_1,  \bar{\mathbf{h}}_2^T \bar{\mathbf{g}}_2, \cdots,  \bar{\mathbf{h}}_c^T \bar{\mathbf{g}}_c) & \\
        & \mathbf{0}_{(K-c) \times (K-c)}
    \end{array}\right]. \label{eq:H_G}
\end{align} 
This implies the properties (\ref{eq:property-1})-(\ref{eq:property-4}).
\qedsymbol

\subsection{Proof of Lemma~\ref{lemma:c}}\label{proof:lemma_c}
First, to determine the value of $c$, observe that
\begin{align}
    \mathrm{rank}(\mathbf{H} \mathbf{G}^T) = 
    \mathrm{rank}(\mathbf{B}_1 \mathbf{H} \mathbf{G}^T \mathbf{B}_2^T) 
     = \mathrm{rank}(\bar{\mathbf{H}} \bar{\mathbf{G}}^T) = c,
\end{align}
where the last equality holds because of (\ref{eq:H_G}).

Now, we design the added matrices in (\ref{eq:designed_M}) as $\mathbf{H}' = \mathbf{B}_1 \bar{\mathbf{H}}'$ and $\mathbf{G}' = \mathbf{B}_2 \bar{\mathbf{G}}'$, where 
\begin{align}
    \bar{\mathbf{H}}' &= \begin{bmatrix}
        \diag(- \bar{\mathbf{h}}_1^T \bar{\mathbf{g}}_1, - \bar{\mathbf{h}}_2^T \bar{\mathbf{g}}_2, \cdots, - \bar{\mathbf{h}}_c^T \bar{\mathbf{g}}_c) \\
        \mathbf{0}_{(K-c) \times c}
    \end{bmatrix}, \\
    \bar{\mathbf{G}}' &= 
    \begin{bmatrix}
        \mathbf{I}_{c} \\
        \mathbf{0}_{(K-c) \times c }
    \end{bmatrix}.
\end{align}
Note that $\mathrm{rank}(\mathbf{M}_o') = 2K$ and 
\begin{align}
    \begin{bmatrix}
        \mathbf{H} & \mathbf{H}'
    \end{bmatrix}
    \begin{bmatrix}
        \mathbf{G} & \mathbf{G}'
    \end{bmatrix}^T
    & = \mathbf{B}_1^{-1} (\bar{\mathbf{H}} \bar{\mathbf{G}}^T + \bar{\mathbf{H}}' (\bar{\mathbf{G}}')^{T}) (\mathbf{B}_2^{-1})^T  \notag \\
    & = \mathbf{0}_{K \times K}.
\end{align}
This proves that the designed $\mathbf{M}_o'$ in (\ref{eq:designed_M}) is an SO matrix.
\qedsymbol

\bibliographystyle{unsrt}
\bibliography{Ref1.bib}

\begin{thebibliography}{10}

\bibitem{shi2021entanglement}
H.~Shi, M.-H. Hsieh, S.~Guha, Z.~Zhang, and Q.~Zhuang.
\newblock Entanglement-assisted capacity regions and protocol designs for quantum multiple-access channels.
\newblock {\em NPJ Quantum Information}, 7(1):74, May 2021.

\bibitem{nielsen2010quantum}
M.~Nielsen and I.~Chuang.
\newblock {\em Quantum Computation and Quantum Information}.
\newblock Cambridge University Press, 2010.

\bibitem{hsieh2008entanglement}
M.-H. Hsieh, I.~Devetak, and A.~Winter.
\newblock Entanglement-assisted capacity of quantum multiple-access channels.
\newblock {\em IEEE Transactions on Information Theory}, 54(7):3078--3090, June 2008.

\bibitem{werner2001all}
R.~Werner.
\newblock All teleportation and dense coding schemes.
\newblock {\em Journal of Physics A: Mathematical and General}, 34(35):7081, August 2001.

\bibitem{song2020capacity}
S.~Song and M.~Hayashi.
\newblock Capacity of quantum private information retrieval with multiple servers.
\newblock {\em IEEE Transactions on Information Theory}, 67(1):452--463, September 2020.

\bibitem{song2021capacity}
S.~Song and M.~Hayashi.
\newblock Capacity of quantum private information retrieval with colluding servers.
\newblock {\em IEEE Transactions on Information Theory}, 67(8):5491--5508, May 2021.

\bibitem{yao_capacity_MAC}
Y.~Yao and S.~Jafar.
\newblock The capacity of classical summation over a quantum {MAC} with arbitrarily distributed inputs and entanglements.
\newblock {\em IEEE Transactions on Information Theory}, 70(9):6350--6370, May 2024.

\bibitem{N_sum_box}
M.~Allaix, Y.~Lu, Y.~Yao, T.~Pllaha, C.~Hollanti, and S.~Jafar.
\newblock {$N$}-sum box: An abstraction for linear computation over many-to-one quantum networks.
\newblock In {\em IEEE GLOBECOM}, pages 5457--5462, February 2023.

\bibitem{aytekin2023quantum}
A.~Aytekin, M.~Nomeir, S.~Vithana, and S.~Ulukus.
\newblock Quantum symmetric private information retrieval with secure storage and eavesdroppers.
\newblock In {\em IEEE GLOBECOM}, pages 1057--1062, December 2023.

\bibitem{dutta2016short}
S.~Dutta, V.~Cadambe, and P.~Grover.
\newblock Short-dot: Computing large linear transforms distributedly using coded short dot products.
\newblock {\em Advances In Neural Information Processing Systems}, 29, July 2016.

\bibitem{ramamoorthy2019universally}
A.~Ramamoorthy, L.~Tang, and P.~Vontobel.
\newblock Universally decodable matrices for distributed matrix-vector multiplication.
\newblock In {\em IEEE ISIT}, pages 1777--1781, July 2019.

\bibitem{das2019distributed}
B.~Das and A.~Ramamoorthy.
\newblock Distributed matrix-vector multiplication: A convolutional coding approach.
\newblock In {\em IEEE ISIT}, pages 3022--3026, July 2019.

\bibitem{yu2017polynomial}
Q.~Yu, M.~Maddah-Ali, and S.~Avestimehr.
\newblock Polynomial codes: an optimal design for high-dimensional coded matrix multiplication.
\newblock {\em Advances in Neural Information Processing Systems}, 30, December 2017.

\bibitem{yao2024inverted}
Y.~Yao and S.~Jafar.
\newblock The inverted 3-sum box: General formulation and quantum information theoretic optimality.
\newblock {\em arXiv preprint arXiv:2407.01498}, 2024.

\bibitem{TIT_catalytic}
T.~Brun, I.~Devetak, and M.-H. Hsieh.
\newblock Catalytic quantum error correction.
\newblock {\em IEEE Transactions on Information Theory}, 60(6):3073--3089, March 2014.

\bibitem{science_brun}
T.~Brun, I.~Devetak, and M.-H. Hsieh.
\newblock Correcting quantum errors with entanglement.
\newblock {\em Science}, 314(5798):436--439, October 2006.

\bibitem{gottesman1997stabilizer}
D.~Gottesman.
\newblock {\em Stabilizer Codes and Quantum Error Correction}.
\newblock California Institute of Technology, 1997.

\bibitem{ketkar2006nonbinary}
A.~Ketkar, A.~Klappenecker, S.~Kumar, and P.~Sarvepalli.
\newblock Nonbinary stabilizer codes over finite fields.
\newblock {\em IEEE Transactions on Information Theory}, 52(11):4892--4914, November 2006.

\bibitem{ashikhmin2001nonbinary}
A.~Ashikhmin and E.~Knill.
\newblock Nonbinary quantum stabilizer codes.
\newblock {\em IEEE Transactions on Information Theory}, 47(7):3065--3072, August 2001.

\bibitem{calderbank1998quantum}
A.~Calderbank, E.~Rains, P.~Shor, and N.~Sloane.
\newblock Quantum error correction via codes over {GF}(4).
\newblock {\em IEEE Transactions on Information Theory}, 44(4):1369--1387, July 1998.

\bibitem{holevo1973bounds}
A.~Holevo.
\newblock Bounds for the quantity of information transmitted by a quantum communication channel.
\newblock {\em Problemy Peredachi Informatsii}, 9(3):3--11, 1973.

\bibitem{newman1997smith}
M.~Newman.
\newblock The {S}mith normal form.
\newblock {\em Linear algebra and its applications}, 254(1-3):367--381, 1997.

\end{thebibliography}

\end{document}